\documentclass[authoryear,11pt]{article}
\usepackage{graphicx}
\usepackage{bbm, bm}
\usepackage[centertags]{amsmath}
\usepackage{amssymb}
\usepackage{amsfonts}
\usepackage{psfrag}

\begin{document}
\title{Mathematical framework of epigenetic DNA methylation in gene body Arabidopsis}
\author{Diana David-Rus $^{*}$\\
\\
$^{*}$Department of Bioinformatics and Structural Biochemistry,\\
 Institute of Biochemistry of Romanian Academy,\\
 \small{Splaiul independentei 296, 060031, Bucharest 17, Romania}}
\date{}
\maketitle
\noindent

\begin{abstract}
In aiming to explain the establishment, maintenance and stability of methylation pattern in gene body of Arabidopsis we propose here a theoretical framework for understanding how the methylated and unmethylated states of cytosine residues are maintained and transmitted during DNA replication. Routed in statistical mechanics, the framework built herein is used to explore minimal models of epigenetic inheritance and identify the necessary conditions for stability of methylated/unmethylated states of cytosine over rounds of DNA replication. The models are flexible enough to allow adding new biological concepts and information. 
\end{abstract}


\section{Introduction}
DNA methylation  is a dynamic epigenetic process that refers to the enzymatic transfer of a methyl group to the specific nucleotides within the DNA
sequence. In eukaryotes, this modification or marks affects almost exclusively
 cytosines [1].
DNA methylation is readily detected in plants
and mammals, where it is critical for normal development
and genome stability. 
Interestingly, plants seem more prone to the inheritance of DNA methylation defects than mammals [2,3]. 

Due to a near-complete genome sequence annotated
to very high standards, a comprehensive set of
genomics tools and powerful genetics, the flowering
plant Arabidopsis has rapidly become a prime model for
the study of DNA methylation and its inheritance
patterns in higher eukaryotes. In this work we will refer at Arabidopsis as a model organism although the framework in which we work it can be applied to DNA methylation in other organisms as well. 

    In the  Arabidopsis, methylation of cytosine has been detected on gene body,  gene promoters and  repeat elements (transposable elements).  If the role of methylation  
in the context of  repeat elements
 is considered to be  of defense against invasive DNA and 
 on gene promoters of silencing the gene,  the role of methylation on gene body is not yet clear.
  On  gene body methylation of cytosine is restricted  to  CG sites, [4,5]  by difference with the repeat elements or gene promoters case where methylation is sequence dependent and can be found also on CHG and CHH sites (where H  can be any of the four nucleotides: A,T,C,G) [6,7]
The relative prevalence of DNA methylation in each
sequence context throughout the genome was assessed,
revealing that 55\% were in CG context, while 23\% and 22\%
were in the CHG and CHH contexts, respectively
[4,5,8,9].

 Gene body methylation occurs on about a third of all genes, and these genes tend to be highly and ubiquitously expressed in different Arabidopsis tissues [10,11]

One of the defining properties of epigenetic phenomena is its stability, the ability of the cell to maintain its epigenetic state stable through many cell divisions. The density of the marks, (methylated states) responsible for the epigenetic effects,  is changed during DNA replication by introducing newly synthesized DNA  indicating that these heritable states must be robust against significant perturbations in there concentration. In the same time, mechanisms of DNA methylation  involve enzymes that can act on more than one nucleotide in its neighborhood. This non-locality of action opens the possibility of interesting collective aspects that have a role in maintaining the  stability of epigenetic states.

 We approach the problem by using methods that traditionally are used in statistical mechanics and dynamical systems.
In my previous work on epigenetic processes [12], I was mostly concerned with understanding  the stability of histone modifications, rather than DNA methylation. 
Given that DNA methylation is another important epigenetic mark critical in development and genome stability, in this work we wish to explore the stability of DNA methylation pattern across multiple generations and focus on understanding DNA methylation in a first approximation in the context of gene body. 

The aim  is to explore the properties of a minimal model of epigenetic silencing in order to identify the necessary conditions for stability of cytosin states that correspond to distinct epigenetic phenotypes i.e. methylated/ un-methylated states. The model is based on the current understanding of DNA methylation in gene body Arabidopsis, with particular emphasis on the interplay between the mechanisms that enable the establishment and maintenance of this modifications [13].

    In section 2  we will present the general framework of the model. 
In order to make the present discussion self-contained we will  introduce  some well known aspects of the methods commonly used in statistical physics (see also [12]). 
In section 3 we will apply  the  framework and  methods presented in previous section to the context of gene body methylation in Arabidopsis and present the results. In  section 4 we will discuss some aspects of the model and in section 5 we will draw the conclusions and present some posible future directions of the model. 

 \section{Methods and  general framework of the model}

We consider a 1D lattice of size $L$  whose sites correspond to nucleotides/cytosines ordered along the length of the DNA.  The nucleotide corresponding to site $i$, can be in several  states, corresponding to particular situation that we are interested in. These states are  labeled by $s=1,...,N$. The rates of transition at site $i$ from state $s'$ to state $s$, namely, $R_{iss'}[s_1,\ldots,s_{i-1},s',s_{i+1},\ldots,s_L]$, depends not only on the local state but also on the states of all the neighbors within a range $l$. In practice, this dependency arises because particular modifications of a site leads to recruitment of particular  enzymes that could affect modification rates of the neighboring nucleotides. 
The master equation describing the time evolution of the probability distribution $P[s_1,\ldots,s_L;t]$ is given by:
\begin{eqnarray}
& &\frac{d}{dt}P[s_1,\ldots,s_L;t]=\nonumber\\
&=&\sum_{i=1}^L \sum_{s'}\big(R_{is_is'}[s_1,\ldots,s_{i-1},s',s_{i+1},\ldots,s_L]P[s_1,\ldots,s_{i-1},s',s_{i+1},\ldots,s_L;t]\nonumber \\
&  &-R_{is's_i}[s_1,\ldots,s_{i-1},s_i,s_{i+1},\ldots,s_L]P[s_1,\ldots,s_{i-1},s_i,s_{i+1},\ldots,s_L;t]\big)
\end{eqnarray}
for times between  DNA replication.
At the point of DNA duplication, a novo strand is formed. This will have as consequences that the fraction of methylated sites right after DNA replication will be diluted. Taking in consideration that right after DNA replication we have a hemimethylated DNA, we represent that in the evolution of probability distribution. Actually for the gene body case, density of methylated sites  will be halved, i.e will be half  the fraction of methylated sites that were before DNA replication. 
 In this process  we assume that DNA duplication happens instantaneously (in reality, fast compared to the time between two duplication events).  
 Fig. 1 provides a schematic representation of the model and its dynamics.
 
\begin{figure}[hbt]
\centering
\includegraphics[width=120mm]{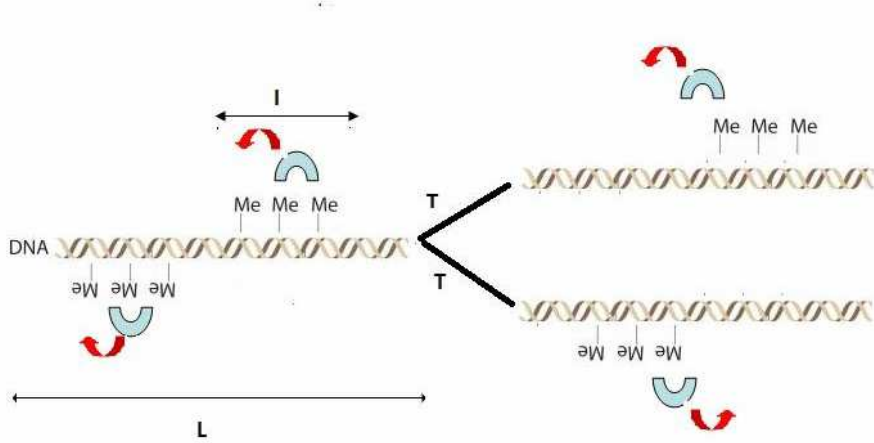}
\caption{\small{Methylation on CG sequence is symetrically on both strands; right after DNA replication at T+ we have a hemimethylated DNA and density of methylated sites is halved; L - length of the lattice; l - range of cooperative behavior; T - periodic time to DNA replication} 
}
\label{fig:1}
\end{figure}

%

To solve the master equation (1) analytically for the long time behavior of $P[s_1,\ldots,s_L;t]$ is generally an impossible task. One, therefore, has to resort to some sort of approximation. One such approximation often used successfully in statistical mechanics is the ``mean field" approximation [14]. In this approach one approximates $P[s_1,\ldots,s_L;t]$ by a factorized form
$\prod_i p_i[s_i;t]$. Using this approximation one derives that the evolution equation for $p_i[s_i;t]$ is going to be
\begin{equation}
\frac{d}{dt}p_i[s_i;t]=\sum_{s'}(\bar{R}_{is_is'}p_i[s';t]-\bar{R}_{is's_i}p_i[s_i;t])
\label{meanfield}
\end{equation}
where the definition of the average rates $\bar{R}_{iss'}$ is
\begin{equation}
\bar{R}_{iss'}=\sum_{s_1,\ldots,s_{i-1},s_{i+1},\ldots,s_L}R_{iss'}[s_1,\ldots,s_L]p_1[s_1;t]\ldots p_{i-1}[s_{i-1};t]p_{i+1}[s_{i+1},t]\ldots p_L[s_L;t].
\end{equation}
Notice that these averaged rates $\bar{R}_{iss'}$ are polynomials in $p_i[s;t]$ making eq.(2) a nonlinear equation.

In the mean field analysis of all the models discussed in this work, we will ignore the spatial variation of `marks' and replace them by average concentrations corresponding to an entire region of DNA, namely $p_i[s_i;t]=p[s_i;t]$. We thereby focus on regions of DNA with one epigenetic fate and be concerned with 'uniform' states. 
The equations for the variables $p[s;t]$ are:
\begin{equation}
\frac{d}{dt}p[s;t]= \sum_{s'}(\bar{R}_{ss'}p[s';t]-\bar{R}_{s's}p[s;t])
\label{meanfieldUniform}
\end{equation}
where  $\bar{R}_{ss'}=\bar{R}_{iss'}$, is given by Eq.~\ref{meanfield}. These are independent of $i$ because the rules of transitions are translation invariant and we  ignore boundary effects.
On incorporating recruitment and cooperative behavior multiple neighboring sites of a site influence the probability of the state at that site, therefore, the transition rates are dependent on what happens on neighboring sites.
We suppose that the rates $R_{is_is'}[s_1, \dots, s_{i-1}, s', s_{i+1}, \dots, s_L]$ depend only on the fraction of sites in a given state in the  neighborhood of $i$ within separation $l$, where $1<<l$ (we could still have $l<<L$ to be physically meaningful). 
We  can group then $L$ sites into $L/l$ clusters of $l$ sites each, i.e. coarse-graining the system. We redefine the probabilities $p_i[s_i, t]$ of state $s_i$ at site $i \in [1,L]$ by the averaged probability $\bar{p}_j[s, t]$ of state $S$ at any cluster $j \in [1, L/l]$, where formally
\begin{equation}
\bar{p}_j[S,t] \equiv \frac{1}{l} \sum_{i=jl-l+1}^{jl} p_i[s_i,t]
\end{equation}
Moreover we can assume that the averaged probabilities are approximately site independent. 
The new states $S$ are not binary corresponding to the presence or absence of marks but a discrete spectrum of states that can be approximated by the concentration of marks in a cluster. This mean-field equivalence of the local probability of a binary state at a site to the probability density (or normalized concentration) of states in a `coarse-grained cluster' is going to be exploited in the rest of the work implicitly in writing down mean-field differential equations for the dynamics of the system. We will not introduce in the rest of the work the formal redefinitions of probabilities done above.
The mean field approximation, turns out, a posteriori, to be justified and quite effective in many cases [14,15]. This method as shown is based on averaged quantities that coarse -grain the system and by neglecting the spontaneous fluctuations in the concentration of the states, predicts long range order.

We will study, analytically and computationally the stochastic model of epigenetic inheritance formulated above for a particular choice of states and rules of state transitions proper for describing DNA methylation in gene body Arabidopsis. In the next section  the discussion will be on a concrete case of DNA methylation in gene body Arabidopsis. Here  we will show  that some restrictions  in the dynamics (transition rates) and some addtional constraints are required for the recovery of the epigenetic marks to take place.

\section{Results: Modeling gene body methylation in Arabidopsis}
 
Abiding by our goal of identifying  minimal models of epigenetic DNA methylation, we develop in this section  a two-state model for studying stable epigenetic marks and understanding  gene body methylation. In Arabidopsis gene body, methylation is restricted almost exclusively to CG sites  and seams to be associated with expression rather then silencing [4,5]. 

Following the understanding of Colot group et. al.[16] for DNA methylation in Arabidopsis 
 we are considering that the process of  methylation  takes places in two critical steps. First step concerns  establishment of DNA methylation pattern and its associated  mechanisms; while the  second concerns  maintenance of this modification  within and between generations.


Based on the general framework presented in previous section  we consider the string of nucleotides  as a 1 dimensional finite lattice that approximate the DNA. As experimentally has been established that methylation status is influenced by nearby cytosine [17] we have to take  into account this fact in generating the model of dynamic evolution of the system.   
Thus, in our model, the rates of transition at site k  depends  of the states of all the neighbors cytosine  within a range l=n, property called cooperativity and of an inherent constant defined by the enzymes involved in the establishing mechanism of DNA methylation. The inherent constant defines the  property of a cytosine to become methylated or de-methylated independently of its neighbors. This can be considered a de novo methylation where a cytosine is methylated/de-methylated with a constant rate due to  enzymatic machinery. By contrast the cooperative term defines the dependency of the state of the cytosine at site k of the methylated status of its n neighbors.
Such a term describes the local modulation  
of the  enzymatic machinery   by raising or lowering the local concentration of enzymes at a given place in the lattice.

In this sense the dynamics that determine the establishment of DNA methylation, the transition rate at site k is defined as bellow by the two components, see fig. 2:  the component that defines the inherent property of cytosine to become methylated or un-methylated independently of its neighbors that we call INHERENT rate and  COOPERATIVITY component which is determined and therefore depends by the state of the n neighbors that surround the cytosine,
\\
\begin{figure}[h]
\centering
\includegraphics[width=150mm]{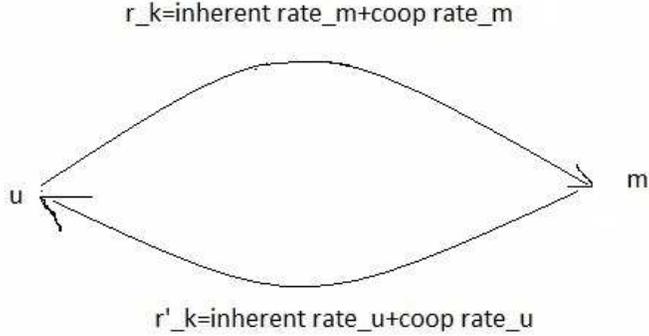}
\label{transition rates} 
\vspace{-5cm}
\caption{\small{transition rates at site k; m-methylated state; u-unmethylated state; $r_k$ is the rate of unmethylated cytosine at site k  to become methylated; $r'_k$  the rate of methylated cytosine at site k to become unmethylated}} 

\end{figure}
Mathematically all of above are written as following:
$r_k=\overbrace{\alpha}^\mathrm{inherent}+\beta\overbrace{\frac{n_k}{2n+1}}^\mathrm{cooperativity};$
$n_k=\sum_{j=k-n}^{k+n}\eta_j(t);$ 
  $\alpha$ is the inherent rate constant of cytosines to be methylated;
 $\beta$ is a proportionality constant;
 $\eta$ is methylated state and $( 1-\eta) $ -de-methylated state.

   $r'_k=\alpha '+\beta '(1-\frac{n_k}{2n+1});$
 $n_k=\sum_{j=k-n}^{k+n}\eta_j(t);$ depends  of  n, number of neighbors;
 where $\alpha'$ is the inherent constant rate of de-methylation ; $\beta'$ is a proportionality constant of de-methylation.




Given the dynamics described above we  write the  equation for  time evolution of the density of methylated sites  for times between  DNA replication known as master equation; with $\langle\eta_k\rangle$ density of methylated sites:


\[\frac{d\langle\eta_k\rangle}{dt}=\langle(1-\eta_k)r_k\rangle-
\langle \eta_k r'_k\rangle;\]
\[\frac{d\langle\eta_k\rangle}{dt}=\langle(1-\eta_k)(\alpha+\beta\frac{\sum_{k-n}^{k+n}\eta_j}{2n+1})\rangle-\langle\eta_k[\alpha'+\beta'(1-\frac{\sum_{k-n}^{k+n}\eta_j}{2n+1})]\rangle;\]
\begin{equation}
\frac{d\langle\eta_k\rangle}{dt}=\alpha+\frac{\beta}{2n+1}\sum_{j=k-1}^{k+n}\langle\eta_j\rangle-\alpha\langle\eta_k\rangle-\frac{\beta}{2n+1}\sum_{j=k-1}^{k+n}\langle\eta_j\eta_k\rangle-(\alpha'+\beta')\langle\eta_k\rangle+\frac{\beta'}{2n+1}\sum_{j=k-1}^{k+n}\langle\eta_j\eta_k\rangle
\end{equation}
To solve analytically  equation (6) we use  mean field approximation mentioned in the previous section, where 
$\langle\eta_k\eta_j\rangle=\langle\eta_k\rangle\langle\eta_j\rangle\equiv P_kP_j, k\neq j$ and also $P_j=P_k=P;$ 
to obtain a simpler equation that describes the dynamics of density of methylated sites at times between DNA replication: 
\[\frac{dP_k}{dt}=\alpha+\frac{\beta}{2n+1}\sum_{j=k-1}^{k+n}P_j-\alpha P_j-\frac{\beta}{2n+1}\sum_{j=k-1}^{k+n}P_jP_k-(\alpha'+\beta')P_k+\frac{\beta}{2n+1}\sum_{j=k-1}^{k+n}P_jP_k;\]
or\[\frac{dP}{dt}=\alpha+\frac{\beta}{2n+1}2nP-\alpha P -\frac{\beta}{2n+1}2nP^2-(\alpha'+\beta')P+\frac{\beta'}{2n+1}2nP^2;\] and by grouping the terms:
\begin{equation}
\frac{dP}{dt}=\alpha+P\underbrace{[\beta\frac{2n}{2n+1}-\alpha-\alpha'-\beta']}_{\Omega}
+\underbrace{[\beta'\frac{2n}{2n+1}-\beta\frac{2n}{2n+1}]}_{\omega}P^2
\end{equation}
In steady state this model  is a quadratic equation therefore  at it's  best can have just one stable state even in the absence of fluctuations induced by perturbations due to DNA replication.

Epigenetic DNA methylation implies alternative states that are stable over time and are inherited through cell division. Any model that tries to explain the methylation process should be able to obtain a coexistence of stable states. Actually the understanding of epigenetic
processes,  in terms of multiple steady
states, has been suggested already long ago by Waddington 
and most clearly by Delbrucks [19,20]
Multistationarity is the property of systems whose structure
is such that they can display two or more distinct 
steady states under identical conditions.
In our model the requirement of mulitistationarity (or  at least of a bistable state)   can happen  if we allow the de-methylated and methylated sites to recruit enzymes cooperatively in a non-linear manner to de-methylated and methylated neighboring sites respectively. 


This will affect the  transition rates for methylation/ de-methylation to include  a degree of non-linear cooperative methylation, respectively de-methylation:  
$r_k=\alpha+\beta\frac{n_k^m}{2n+1}$;  $r'_k=\alpha '+\beta '(1-\frac{n_k^p}{2n+1}),$ 
$n_k=\sum_{j=k-n}^{k+n}\eta_j(t)$ where m is the degree of non-linear cooperative methylation and  p is the degree of non-linear cooperative de-methylation.  

In this new context, the quadratic  equation (7) is now changed  to   a polynomial of higher order and the model is modified to enable the presence of multiple dynamical attractors. Even for the simplest case of non-linear cooperative behavior when m=p=2 we get in steady state a polynomial of order 3 instead of the quadratic equation (7) (see Appendix A), obtaining  a model that can have a bi-stable state and as such the requirement for epigentic DNA methylation and memory. 

 In fact  if we are calling f(a) the right side of such polynomial equation for m=p=2, f(a) will have three zeros,
$a_1< a_2< a_3$ in the interval (0,1). The scenario relevant to us is when $a_1$ and $a_3$ are stable and are separated by $a_2$ unstable ($a_3$ corresponds to high concentration of marks and $a_1$ to low concentration of marks) see fig. 3 
\begin{figure}[h]
\centering
\includegraphics[width=90mm]{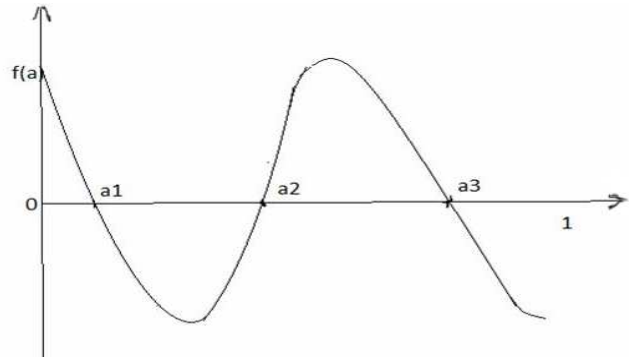}
\label{polynomial}
\caption{ polynomial equation of order 3 when m=p=2}

\end{figure}

Any initial states with $a(0)< a_2$ will eventually be attracted to $a_1$ while any initial state with $a(0)> a_2$ will  be attracted  to $a_3$. Suppose now that DNA replication takes place  periodically at time T. Right after replication the density of methylated sites is halved (see fig.1). It is clear then that if $a_2\geq \frac{1}{2}a_3$, there will be  only one stable fixed point, which will be close to $a_1$ if T is large enough (time T to replication considerably larger then the time scale of methylation rates).
To see this we simply note that starting from $a_3$ the cell after DNA replication will have a value which is less than $a_2$ and so will enter the basis of atraction of the stable fixed point $a_1$.

 However for  $a_2 < \frac{a_3}{2}$ and T fulfilling the same conditioned stated earlier, then there will be two stable fixed point, one near $a_1$ and one near $a_3$ even with perturbations induced by DNA replication. These puts restrictions on the parameters entering f(a) and T required for the existence of multiple stable points and thus of epigenetic memory and regulation; the density of methylated sites right after DNA replication (after halving) has to have a higher value then $a_2$ the unstable value. 
 We shall not go into this here.
 It is also possible to give fairly explicit expressions for T in terms of f(a), (see Appendix B)

Going beyound mean field, using Monte Carlo simulations 
\begin{figure}[h]
\centering
\includegraphics[width=100mm]{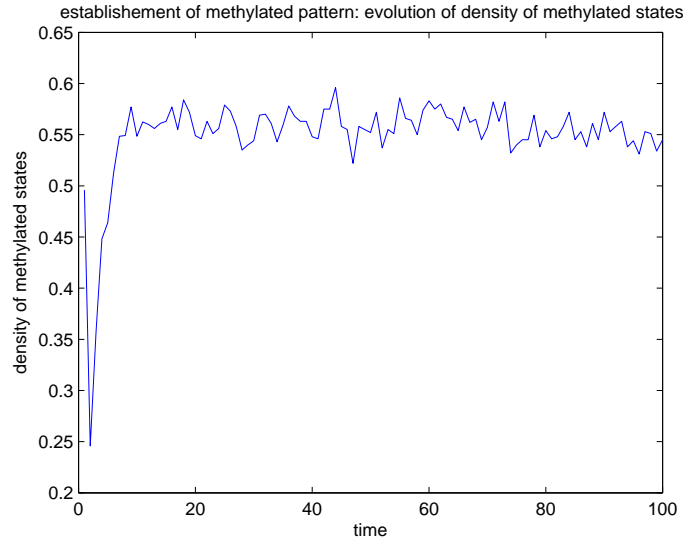}
\label{establishing}
\caption{\small{ Establishment of methylated pattern in the absence of DNA replication- evolution of the density of methylated states; 1unit time $\approx$ 1 mcs.}}
\end{figure}
we would like to see if the requirement of stability is mainteined in the above framework  when the perturbations induced by DNA replications are taken in consideration.  
In this sense fig 4. shows  how methylation pattern is established and remain stable, reaching a stationary state from following the dynamics introduced so far. The DNA replication  is not yet involved  in the process.

 When we introduced in  the system the perturbation due to DNA replication by simply halving periodically the density of methylated sites   we see in fig. 5 that the methylated pattern, the density of methylated sites is recovered, provided that the time to replication is longer then the recovery time, showing as such, explicitly that the dynamics described suffix for the stability of the methylated pattern.
 \begin{figure}[h,t]
\centering
\includegraphics[width=90mm]{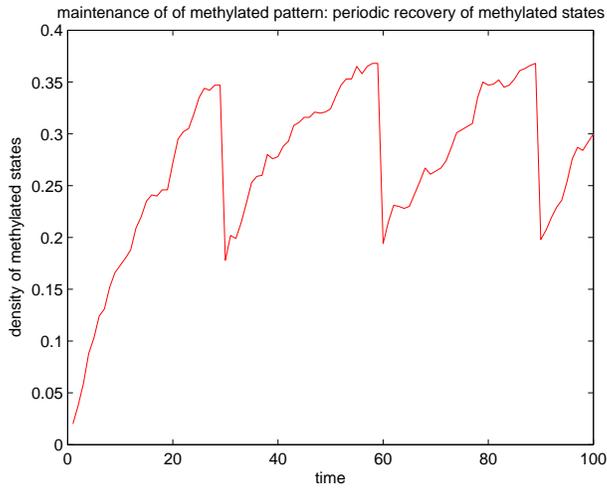}
\label{fig:5}
\caption{\small{maintenance  of methylated pattern in the presence of DNA replication-periodic recovery of methylated states; 1unit time $\approx$ 1 mcs.} }
\end{figure}

 After multiple replication cycles one expects methylation to not be exactly as it was originally because of 'accidental' loss or gain, see fig. 5. For  Monte Carlo simulation details and how the system was prepared see Appendix C. All the parameters used in the simulation were chosen based on  literature research.





\section{Discussions}

The requirement of non-linearity  to can have multiple dynamical attractors in the system has been previously studied in other epigenetic contexts [2l]. This studies showed  that to obtain multistationarity in the system a positive feedback circuit is a necessary condition. We note, however, that at a fundamental level the presence of non-linear cooperative recruitment of enzymes to methylated/de-methylated states  might be due to the existence  of such positive feedback circuit. 
 In Arabidopsis more states are introduced when methylation takes place  on repeat elements over their entire length and strongly correlates with transcription inhibition. Methylation in this case is on CG, CHG and CHH sites instead of solely CG sites like on gene body, and as such is sequence dependent. 
We know from previous work [12] that the presence of intermediate
states naturally lead to cooperative effects when each of the
intermediate states recruit enzymes for further modification. A continuation of this work would be to analyse the effects of sequene dependency on the stability of the methylation pattern and try to understand if non-linearity in the transition rates is still a key ingredient in the maintenance of DNA methylation  or if  solely the presence of more states is stabilizing the system.
 We have phrased the mean-field theory in terms of coarse grained
quantities like the fraction of sites with a particular mark
in a cluster. Given that methylation on repeat element takes place over their entire length, a
natural question would be to try understand how does the effective model change if we
continue the coarse-graining to larger length scales. In other
words: one could ask how the model 'renormalizes' under
iterative blocking transformations [14,15]. In absence of any conservation
law, there is no obvious reason why this system should not have a
finite (although long) correlation length in space and, similarly,
a finite correlation time. The system would not have genuinely
multiple phases. All these effects, which are missed by mean-field
theory, would, in principle, show up in renormalization group.

\section{ Conclusions}
The stability of epigenetic DNA methylation is a rich subject in biology. The exact function(s), of much of the DNA methylation found outside of repeat elements remain unclear [6,7].  Using an approach routed in statistical physics we  proposed a theoretical framework for understanding how the methylated and non-methylated states of cytosine are maintained and transmitted when perturbations (as DNA replication) are involved.

The model explains
	the establishment of DNA methylation pattern i.e. study the dynamics of the density of the methylated sites. Analyzes the effect of the non-linear cooperativity (in the transition rates) on the  stability of the marks and shows that at least in the gene body case where methylation is restricted to CG sites in order to have stable DNA methylation patterns transmitted over generations non-linear cooperativity is required in the maintenance process.

Many features presented here in the context of Arbidopsis can be also extendet to DNA methylation in other organisms as well.  The model extends the view that multi-stationarity in gene body DNA methylation pattern arises by allowing the de-methylated and methylated sites to recruit enzymes cooperatively in a non-linear manner to de-methylated and methylated neighboring sites respectively. We finish by concluding that overall the present work  lays the grounds for understanding DNA methylation on repeat elements in Arabidopsis, and as such extending the mathematical framework  to compleate the modelling and understanding of DNA methylation in Arabidopsis. 
\section*{Acknoledgements}
I would like to add special thanks to Prof. Joel L.Lebowitz for his valuable insights. I also acknowledge useful discussions with  Prof. Andrei J. Petrescu and Prof. Vincent Colot. This work was supported by  Fondation Pierre-Gilles de Gennes-postdoctoral fellowship and POSDRU/89/1.5/S/60746.21

\section*{References}

[1] A. Bird, Genes $\&$ Dev. 16, 6 (2002).

[2] Richards, E. J. I, Nat. Rev. Genet. 7, 395 (2006)

[3] Whitelaw, N. C. and E. Whitelaw Curr Opin Genet Dev,18,273(2008)

[4] Shawn J. Cokus, et. all, Nature. March 13; 452(7184), 215 (2008)

[5] Lister R, O'Malley RC, et.all, Cell 2008;133,523

[6] Henderson IR, Jacobsen SE., Nature. May 24;447(7143),418 (2007)

[7] Suzuki MM, Bird A. Nat Rev Genet. Jun9, 6:465 (2008)

[8] Zhang X, Yazaki J, Sundaresan A, et al,Cell,126,1189 (2006)

[9] Zilberman D, Gehring M, Tran RK, Ballinger T, Henikoff S., Nat Genet,39,61 (2007)

[10] Zilberman D, Henikoff S, Development, 134, 3959 (2007)

[11] Zhang X, Yazaki J, Sundaresan A, Cokus S, Chan SW, et al. Cell, , 126,1189 (2006)

[12] Diana David-Rus, Joel L. Lebowitz , Journal of Theoretical Biology, 258, 112(2009)

[13] Teixeira FK, Colot V. EMBO J,28, 997 (2009)

[14]  Linda E. Reichl, A Modern Course in Statistical Physics , Willey Verlag (2009)

[15] David Chandler, Statistical mechanics, Oxford University Press, (1987)

[16] Felipe Karam Teixeira, Vincent Colot,The EMBO Journal, 28, 997 (2009)

[17] Zhang X, Shiu S, Shiu S, Cal A, Borevitz JO,. PLoS Genet,4, (2008)

[18] Zilberman D, Gehring M, Tran RK, Ballinger T, Henikoff S., Nat Genet, 39 (2007)

[19] C. H. Waddington, Symp. Soc. Exp. Biol.,2, 145(1949)

[20] Delbruck, Colloq. Int. C. N. R. S.,8, 33(1949)

[21] R. Thomas and M. Kaufman, Chaos, 11, 170 (2001)

[22]  J. Demongeot, J. Biol. Syst., 6,1(1998)

 \end{document}


\section*{Appendix}
\appendix
\section{}

\[\frac{d\langle\eta_k\rangle}{dt}=\langle(1-\eta_k)r_k\rangle-\langle\eta_kr_{k}'\rangle ;\]
replacing $r_k=\alpha+\beta\frac{n_k^m}{2n+1}$;  $r'_k=\alpha '+\beta '(1-\frac{n_k^p}{2n+1}),$ for m=p=2  in the above eq. we obtain:

\[\frac{d\langle\eta_k\rangle}{dt}=\alpha+\frac{\beta}{2n+1}\Big(\sum_{j=k-n}^{k+n}\langle\eta_j\rangle\Big)^2-\alpha\langle\eta_k\rangle-\frac{\beta}{2n+1}\langle\eta_k\rangle\Big(\sum_{j=k-n}^{k+n}\langle\eta_j\rangle\Big)^2-\]
\[-\alpha'\langle\eta_k\rangle\ -\beta'\langle\eta_k\rangle+\frac{\beta'}{2n+1}\langle\eta_k\rangle\Big(\sum_{j=k-n}^{k+n}\rangle\eta_j\rangle\Big)^2\]
Using mean field approximation: $\langle\eta_k\eta_j\rangle=\langle\eta_k\rangle\langle\eta_j\rangle\equiv$ $P_kP_j=P^2;$ and 

$\sum_{j=k-n}^{k+n}P_j=2nP$
we get: 
\[\frac{dP_k}{dt}=\alpha+\frac{\beta}{2n+1}\Big(\sum_{j=k-n}^{k+n}P_j\Big)^2-\alpha P_k-\frac{\beta}{2n+1}P_k\Big(\sum_{j=k-n}^{k+n}P_j\Big)^2-\alpha' P_k -\beta' P_k +\frac{\beta'}{2n+1} P_k\Big(\sum_{j=k-n}^{k+n}P_j\Big)^2\]
and given that $P_j=P_k=P$ we obtain the polynomial of 3rd order:

\[\frac{dP}{dt}=\alpha -P(\alpha+\alpha'+\beta')+ \frac{4n^2\beta}{2n+1}P^2+4n^2\Big(\frac{\beta'-\beta}{2n+1}\Big)P^3\]

\section{ }
Let $0\leq a(t)\leq 1$ be the fraction of marked sites. In the mean field description we can formally define:
\[\frac{da(t)}{dt}=\Pi(a_i-a(t))=f(a)\]

where $0\leq a_1 <a_2<a_3...<a_{2k+1} \leq 1$.

We choose an odd number of stationary points since we want $f(0)\geq 0$, $f(1)\leq 0$. The odd zeros of f(a), $a_1,a_3...a_{2k+1}$ will be linearly stable fixed points while the even number roots will be unstable fixed points. 

If we consider now the effect of DNA replication  when the fraction of methylated sites is halved then the new "fixed points" corresponding to the stable fix points $a_j$ will have a fraction of methylated sites right after mitosis
$a_i^*$ with $a_0=0<a_1^*<(1/2)a_1,....a_{2j}<a_{2j+1}^*<(1/2)a_{2j+1},...$.

Let $T_i$ be the period to DNA replication in which the fraction of marked sites will increase from $a_j^*$ to $2a_j^*$during one cycle then integrating (9), we get 

\[T_i=\int_{a_i^*}^{2a_i^*} \frac{ds}{\Pi(a_i-s)}=\sum_{j=1}^{2k+1}B_j log\frac{a_j-a_i^*}{a_j-2a_i^*}\]
where the $B_j$ can be computed in terms of the ${a_i}$. \\

\section{}
In preparing the system we generate a 1D lattice of L sites (0's and 1's) with periodic bounday conditions. For our simulation we used L=1000. Each site is a   CG nucleotides that has the potential of becoming methylated or dimethylated. 
We defined two probability distributions: $P_1$ is the probability of site i to be methylated, $P_0$ is the probability of site i to be dimethylated. Each probability distribution is constructed based on the transition rates of being methylated or dimethylated described in the paper. The number of neighbors n around each site i that we randomly pick is kept fixed, and we sum the number of methylated sites over this neighbors. In the simulation we used n=50. In one iteration we've done as following: we start with an initial random configuration of sites being methylated and dimethylated. 
We pick up a site i at random, if $\sum_{j=i-n}^{i+n}x(j)\leq P_1$ keep the site methylated, else revert to dimethylated; and if $\sum_{j=i-n}^{i+n}x(j)\leq P_0$ keep the site dimethylated, else revert to being methylated. 
I'm doing this L times, each time based on previous configuration.  I calculate then  the number of methylated sites and normalize to the length of the lattice. This will give me the density of methylated sites at time t.  In my simulation one unit of time is echivalent with one monte carlo step. 
To simulate the DNA replication process, we are introducing a periodic fluctuation that has as effect the halving of the density of metylated sites periodically at time T. In the simulation T is echivalent with 30 monte carlo steps.